# FINITE SIZE INDUCED PHENOMENA IN 2D CLASSICAL SPIN MODELS

*O. Kapikranian[1], B. Berche[2], Yu. Holovatch[3]*

[1]*Institute for Condensed Matter Physics, National Acad. Sci. of Ukraine, 79011 Lviv, Ukraine*
*Laboratoire de Physique des Matériaux, Université Henri Poincaré, Nancy 1, 54506*
*Vandœuvre les Nancy Cedex, France*
e-mail: akap@icmp.lviv.ua

[2]*Laboratoire de Physique des Matériaux, Université Henri Poincaré, Nancy 1, 54506*
*Vandœuvre les Nancy Cedex, France*
e-mail: berche@lpm.u-nancy.fr

[3]*Institute for Condensed Matter Physics, National Acad. Sci. of Ukraine, 79011 Lviv, Ukraine*
*Institut für Theoretische Physik, Johannes Kepler Universität Linz, 4040 Linz, Austria*
e-mail: hol@icmp.lviv.ua

We make a short overview of the recent analytic and numerical studies of the classical two-dimensional XY and Heisenberg models at low temperatures. Special attention is being paid to an influence of finite system size $L$ on the peculiarities of the low-temperature phase. In accordance with the Mermin-Wagner-Hohenberg theorem, spontaneous magnetisation does not appear in the above models at infinite $L$. However it emerges for the finite system sizes and leads to new features of the low-temperature behaviour.

Celebrated Mermin-Wagner-Hohenberg and Bogolyubov's $1/q^2$ theorems forbid spontaneous symmetry breaking in two-dimensional (2D) systems of continuous symmetry [1]. Therefore, there is no spontaneous magnetisation in the 2D XY and Heisenberg models for temperatures $T \neq 0$. This well–known fact does not concern the above models on lattices of finite size $L$. In applied research this can concern the case of thin ferromagnetic films (always finite in practice) [2] and some other objects. On the other hand models of finite size are highly important for analysing Monte Carlo simulations which are restricted to finite lattices as well.

The presence of residual magnetisation in the finite XY and Heisenberg models in 2D can be simply argued by the obvious fact that in finite systems transition to the ordered phase at $T=0$ (when all spins of the XY and Heisenberg models are pointed in the same direction) cannot be discontinuous. But besides this trivial prediction other interesting phenomena can occure like the quasi-long-range ordering (QLRO) in the 2D XY model [3]. A challenging question is whether QLRO is possible in the Heisenberg model in two dimensions or not. Although it is commonly believed that it is not possible due to the absence of stable topological defects, there are still some contrary arguments [4].

Here we summarize resent analytic and Monte Carlo results for the XY and Heisenberg models on finite two-dimensional lattices. The analytic approach is based on the spin-wave approximation (SWA) which is known to give a good estimation of the exact characteristics of the 2D XY model at low temperatures [5]. Although the reliability of the SWA for the two-dimensional Heisenberg model of infinite size is not proven we assume that it holds for a finite lattice. We pay attention to the magnetisation as a function of lattice size and to the spin-spin correlation function as a function of distance between sites. In order to support the analytic treatment we present results of Monte Carlo simulations of the XY and Heisenberg models on two dimensional lattices of different sizes that we have performed for different values of temperature. Moreover we show the results of our Monte Carlo similations of the 2D XY model with quenched structural dilution.

The generalized Hamiltonian for both the XY and Heisenberg models can be written as

$$H = -J \sum_{<r,r'>} \vec{S}_{\vec{r}} \vec{S}_{\vec{r}'} , \qquad (1)$$

where the summation is over the nearest neighbour sites of a 2D square lattice of linear size $L$, $J>0$ is the interaction constant, with $\vec{S}_{\vec{r}} \vec{S}_{\vec{r}'} = S^x_{\vec{r}} S^x_{\vec{r}'} + S^y_{\vec{r}} S^y_{\vec{r}'}$ for the XY model and for the Heisenberg model, $\vec{S}_{\vec{r}} \vec{S}_{\vec{r}'} = S^x_{\vec{r}} S^x_{\vec{r}'} + S^y_{\vec{r}} S^y_{\vec{r}'} + S^z_{\vec{r}} S^z_{\vec{r}'}$. The low-temperature behaviour is deeply connected to the symmetry of the models. In the 2D XY model rotations of spins form an Abelian group that allows formation of stable topological defects like spin vortices. In this sense it can be called an Abelian model in contrast to non-Abelian ones. The 2D Heisenberg model is non-Abelian and therefore there are no stable topological defects and as a consequence no QLRO in the infinite system.

The quantities we pay attention to in the present work are the magnetisation *M*:

$$M = \frac{1}{N}\sum_{\vec{r}}\left\langle \left|\vec{S}_{\vec{r}}\right|\right\rangle = \left\langle \left|\vec{S}_0\right|\right\rangle, \quad (2)$$

and the spin-spin correlation function $G_2(R)$:

$$G_2(R) = \left\langle \vec{S}_{\vec{r}}\vec{S}_{\vec{r}+\vec{R}}\right\rangle. \quad (3)$$

Introducing angle variables $\vartheta_{\vec{r}}$: $S_{\vec{r}}^x = \cos\vartheta_{\vec{r}}$, $S_{\vec{r}}^y = \sin\vartheta_{\vec{r}}$, for the *XY* model we can write the scalar product of spins as

$$\vec{S}_{\vec{r}}\vec{S}_{\vec{r}'} = \cos(\vartheta_{\vec{r}} - \vartheta_{\vec{r}'}), \quad (4)$$

while spin of the Heisenberg model has two degrees of freedom, so we have two angles $\vartheta_{\vec{r}}^{(1)}$ and $\vartheta_{\vec{r}}^{(2)}$: $S_{\vec{r}}^x = \cos\vartheta_{\vec{r}}^{(1)}\cos\vartheta_{\vec{r}}^{(2)}$, $S_{\vec{r}}^y = \sin\vartheta_{\vec{r}}^{(1)}\cos\vartheta_{\vec{r}}^{(2)}$, $S_{\vec{r}}^z = \sin\vartheta_{\vec{r}}^{(2)}$. Then a scalar product of two Heisenberg spins reads:

$$\vec{S}_{\vec{r}}\vec{S}_{\vec{r}'} = \cos(\vartheta_{\vec{r}}^{(1)} - \vartheta_{\vec{r}'}^{(1)})\cos(\vartheta_{\vec{r}}^{(2)} - \vartheta_{\vec{r}'}^{(2)}) + (1-\cos(\vartheta_{\vec{r}}^{(1)} - \vartheta_{\vec{r}'}^{(1)}))\sin\vartheta_{\vec{r}}^{(2)}\sin\vartheta_{\vec{r}'}^{(2)}. \quad (5)$$

Since our work concerns low temperature properties, we can assume all the spins of the *XY* and Heisenberg models being pointed approximately in the same direction. This assumption means that all angles $\vartheta_{\vec{r}}$ and $\vartheta_{\vec{r}}^{(1)}$, $\vartheta_{\vec{r}}^{(2)}$ stay small. Therefore the spin-wave approximation (SWA) can be applied, i. e. we can expand the trigonometric functions in (4) and (5) keeping quadratic terms only. Then the Hamiltonian of the 2*D XY* model can be written as

$$H^{XY} = H_0 + H^{XY}(\vartheta) \quad (6)$$

with

$$H^{XY}(\vartheta) = \frac{1}{2}J\sum_{<\vec{r},\vec{r}'>}(\vartheta_{\vec{r}} - \vartheta_{\vec{r}'})^2. \quad (7)$$

The Hamiltonian of the Heisenberg model in the SWA can be expressed through (7) as well:

$$H^{Heis} = H_0 + H^{XY}(\vartheta^{(1)}) + H^{XY}(\vartheta^{(2)}). \quad (8)$$

It has been proved that the SWA gives a quite nice estimation to the true low-temperature behaviour of the 2*D XY* model in the thermodynamic limit. Although the behaviour of the infinite 2*D* Heisenberg model at low temperatures is believed to be quite different, we can assume that for finite lattices the SWA will give reliable result for the low-temperature characteristics of the model.

In the case of the 2*D XY* model, the asymptotic behaviour of the spin-spin correlation function is well known [5]:

$$G_2^{XY}(R) = \left\langle \cos(\vartheta_{\vec{r}} - \vartheta_{\vec{r}+\vec{R}})\right\rangle \propto R^{-\eta^{XY}} \quad (9)$$

with $\eta^{XY} = kT/(2\pi J)$. The finiteness of the lattice causes a neglegible correction to the exponent $\eta^{XY}$. The magnetisation in the model on a finite lattice decays with the linear size $L$ according to a power law [6]:

$$M^{XY} = L^{-\eta^{XY}/2} \quad (10)$$

with the same exponent $\eta^{XY}$ that stands in (9).

For the 2*D* Heisenberg model, the separation of coordinates $\vartheta_{\vec{r}}^{(1)}$ and $\vartheta_{\vec{r}}^{(2)}$ in (8) allows us to rewrite the spin-spin correlation function through the correlation function of the *XY* model:

$$G_2^{Heis}(R) = \left[G_2^{XY}(R)\right]^2. \quad (11)$$

Then we have

$$G_2^{Heis}(R) \propto R^{-\eta^{Heis}} \quad (12)$$

with $\eta^{Heis} = 2\eta^{XY}$. It is important to stress here that this result can be obtained analytically only if the temperature is taken in the limit $kT \to 0$ but the lattice remains finite. This is due to the fact that $\vartheta_{\vec{r}}^{(1)}$ and $\vartheta_{\vec{r}}^{(2)}$ are not independent variables, but can be considered as independent only when the temperature approaches zero [7].

A similar outcome can be obtained for the magnetisation of the 2*D XY* model as a function of the linear lattice size $L$:

$$M^{Heis} \propto L^{-\eta^{Heis}/2}, \quad (13)$$

with the same $\eta^{Heis}$ that stands in the correlation function (12).

To check the above analytic results we have performed Monte Carlo simulations of the Heisenberg spin model on lattices of different sizes and at different temperatures. The Wolff's cluster algorithm was used for this purpose [8]. The exponent $\eta$ (Fig.1) was obtained on the base of three different observables, analysing the finite size scaling of the magnetisation, $M \propto L^{-\eta(T)/2}$, the pair correlation function, $G_2(L/2) \propto L^{-\eta(T)}$, and the magnetic susceptibility, $\chi \propto L^{2-\eta(T)}$. All three quantities were computed at different temperatures for varying system sizes, giving access to a temperature dependent exponent $\eta(T)$. Power-law scaling found for all three quantities $M$, $G_2$ and $\chi$ supports the presence of a QLRO phase found by analytic considerations.

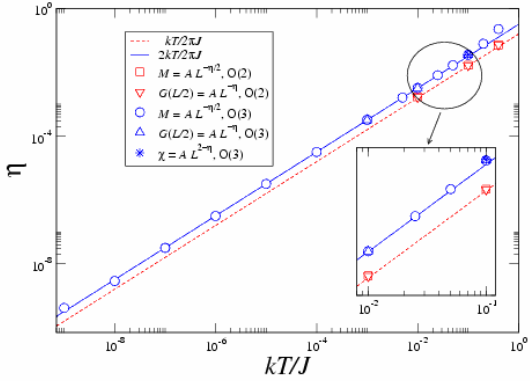

*Fig.1. Comparison between the exponents of the two-dimensional Heisenberg model, $\eta^{Heis}$, obtained from the Monte Carlo simulations and from the analytic calculation in the SWA. The dashed line presents $\eta^{XY}$.*

Moreover we have performed Monte Carlo simulations of 2D *XY*-spins on regular and diluted lattices [9] using the same Wolff's cluster algorithm. According to Harris criterion [10], disorder is irrelevant at the BKT transition of the two-dimensional *XY* model. As a consequence, the universality class is unchanged ($\eta = ¼$) at the transition but randomness has a strong influence at low temperature in the critical phase of the model. To study this effect numerically, we have to average the physical quantities over many realizations of disorder. For each realization, we discarded typically $10^5$ sweeps for thermalization, and the measurements were performed with typically $10^5$ production sweeps. Disorder averages were then performed using typically $10^3$ samples. The boundary conditions were chosen periodic.

The quantity we call magnetisation is the thermodynamic average of the real instant magnetisation which varies, for a given realization of disorder, from one MC iteration to the next. There are reasons to investigate the distribution of instant magnetisation, since it can give some information about the inner nature of the model. A convenient way to display this distribution is to draw a ring function which can be defined in the following way: it is obtained when one plots the successive values of the magnetisation (for each Monte Carlo step) in the plane ($m_x$, $m_y$) where $m_x$ and $m_y$ are the two components of magnetisation (Fig.2).

The shape of the ring functions shows clear non-Gaussian character as the temperature increases which lies in the fact that more points are situated in the inner region of the rings. As it should be, the mean magnetisation tends to zero with increase of temperature in both pure and diluted systems. The "diluted" ring functions are always smaller, since we consider magnetisation per site taking all sites into account, even those which are vacant.

The finite size scaling of the magnetisation, Eq.(10), can be deduced from the study of the ring functions for different sizes of the lattice. Thus we obtain the Monte Carlo result for the exponent $\eta^{XY}_{dil}$ of the pair correlation function of a diluted system. In this case it will depend on concentration of dilution. Again we have nice agreement between Monte Carlo and the perturbation expansion for the spin-wave approximated Hamiltonian which gives [11]:

$$\eta^{XY}_{dil} = \eta^{XY}\left(1/c^2 + 0.73(1-c)/c^3 - 2.27(1-3c+2c^2)/c^4\right),$$

where $c$ is the concentration of magnetic sites.

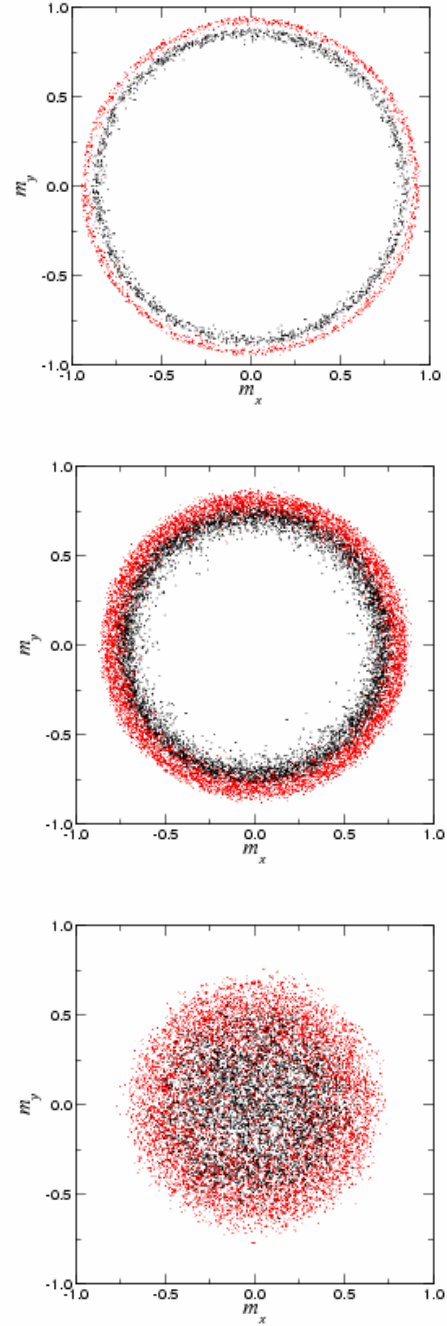

*Fig.2. Ring functions for a system of size L=16 at dilution c= 0.95 for temperatures (from top to bottom) kT/J= 0.3, 0.7 and 1.1. The outer ring function (red on-line) represents the pure system.*

We have presented both analytic and Monte Carlo approach to investigate finite two-dimensional spin models of continuous symmetry. The region of applicability of our analytic treatment is restricted to low temperatures only, since we work in the SWA.

The results of the Monte Carlo simulations are found in good agreement with the analytic results in a wide range of low temperatures.

We continue to work in this direction applying structural disorder to the Heisenberg model in two dimensions and introducing also correlated impurities in both models. The potential goal of this kind of research is to discover the exact nature of interaction of non-magnetic impurities with topological defects and as a consequence their influence on the QLRO phase characteristics. While some steps in this direction in Monte Carlo simulations can be found in literature, a reliable analytic approach to the problem is not developed up to our knowledge.

This work was supported by the CNRS-NAS exchange program.

**REFERENCES**

1. N. D. Mermin, H. Wagner. Absence of ferromagnetism or antiferromagnetism in one- or two-dimensional isotropic heisenberg models, Phys. Rev. Lett.. **22** (1966) 1133; N.N. Bogolyubov, *Selected works*, **3**, Kiev (1970) (In Russian)
2. S. T. Bramwell and P. C. W. Holsworth. Magnetization and universal sub-critical behaviour in two-dimensional *XY* magnets. J. Phys.: Condens. Matter **5** (1993) L53
3. J. M. Kosterlitz, D. J. Thouless. Ordering, metastability and phase transitions in two-dimensional systems. J. Phys. C: Solid State Phys. **6** (1973) 1181
4. A. Patrascioiu, E. Seiler. The difference between Abelian and non-Abelian models: fact and fancy. Preprint math-ph/9903038 (1999)
5. F. Wegner. Spin-ordering in a planar classical Heisenberg model. Z. fur Phys. **206** (1967) 465
6. J. Tobochnik, G. V. Chester. Monte Carlo study of the planar spin model. Phys. Rev. B **20** (1979) 3761; P. Archambault, S. T. Bramwell, P. C. W. Holdsworth. Magnetic fluctuations in a finite two-dimensional *XY* model. J. Phys. A: Math. Gen. **30** (1997) 8363-8378; S. T. Bramwell et al. Magnetic fluctuations in the classical *XY* model: The origin of an exponential tail in a complex system. Phys. Rev. E **63** (2001) 041106
7. O. Kapikranian, B. Berche, Yu. Holovatch. Quasi-long-range ordering in a finite-size 2*D* Heisenberg model. Preprint hep-th/0611264 (submitted to J. Phys. A)
8. U. Wolff. Collective Monte Carlo updating for spin systems. Phys. Rev. Lett. **62** (1989) 361
9. O. Kapikranian, B. Berche, Yu. Holovatch. The 2*D XY* model on a finite lattice with structural disorder: quasi-long-range ordering under realistic conditions. Preprint cond-mat/0612147 (submitted to Eur. Phys. Jour. B)
10. A. B. Harris. Effect of random defects on the critical behaviour of Ising models. J. Phys. C: Solid State Phys. **7** (1974) 1671
11. O. Kapikranian, B. Berche, Yu. Holovatch. Perturbation expansion for the diluted two-dimensional *XY* model. Preprint cond-mat/0611712 (submitted to Phys. Lett. A)

**ЭФФЕКТЫ КОНЕЧНОГО РАЗМЕРА В ДВУХМЕРНЫХ КЛАССИЧЕСКИХ МОДЕЛЯХ**

*О. Капикранян, Б. Берш, Ю. Головач*

Приведен краткий обзор недавних аналитических и численных исследований классической XY модели и модели Гейзенберга при низких температурах. Специальное внимание уделено влиянию конечного размера системы $L$ на свойства низкотемпературной фазы. Согласно теоремы Мермина-Вагнера-Хогенберга спонтанная намагниченность отсутствует в этих моделях при бесконечном $L$. Однако она появляется в системах конечного размера и приводит к новым чертам низкотемпературного поведения.

**ЕФЕКТИ СКІНЧЕНОГО РОЗМІРУ В ДВОВИМІРНИХ КЛАСИЧНИХ МОДЕЛЯХ**

*О. Капікранян, Б. Берш, Ю. Головач*

Зроблено короткий огляд недавніх аналітичних і числових досліджень класичної XY моделі і моделі Гайзенберга при низьких температурах. Особливу увагу приділено впливу скінченого розміру системи $L$ на властивості низькотемпературної фази. За теоремою Мерміна-Вагнера-Гогенберга спонтанна намагніченість в цих моделях відсутня при безмежному $L$. Проте вона з'являється в системах скінченого розміру і спричиняє нові риси низькотемпературної поведінки.